\documentclass[pra,twocolumn,a4paper,showpacs,superscriptaddress]{revtex4}
\usepackage{amsmath}
\usepackage{amsfonts}
\usepackage{graphicx}
\usepackage{longtable}
\newcommand{\be}{\begin{equation}}
\newcommand{\ee}{\end{equation}}

\newcommand{\ba}[1]{\left(\begin{array}{#1}}
\newcommand{\ea}{\end{array}\right)}
\begin{document}

\title{Interplay of quantum stochastic and  dynamical maps to discern  Markovian and non-Markovian transitions} 
\author{A. R. Usha Devi}
\email{arutth@rediffmail.com}
\affiliation{Department of Physics, Bangalore University, 
Bangalore-560 056, India}
\affiliation{Inspire Institute Inc., Alexandria, Virginia, 22303, USA.}
\author{A. K. Rajagopal} 
\affiliation{Inspire Institute Inc., Alexandria, Virginia, 22303, USA.}
\author{Sudha} 
\affiliation{Inspire Institute Inc., Alexandria, Virginia, 22303, USA.}
\affiliation{Department of Physics, Kuvempu University, Shankaraghatta, Shimoga-577 451, India.}
\author{R. W. Rendell} 
\affiliation{Inspire Institute Inc., Alexandria, Virginia, 22303, USA.}
\date{\today}

\begin{abstract} 
It is known that the dynamical evolution of a system, from an initial tensor product state of system and environment, to any two later times, $t_1,t_2\ (t_2>t_1)$, are both completely positive (CP) but in the intermediate times between $t_1$ and $t_2$ it need not be CP. This reveals the key to the Markov (if CP) and nonMarkov (if it is not CP) avataras of the intermediate dynamics. This is brought out here in terms of the  quantum stochastic map $A$ and the associated dynamical map $B$ -- without resorting to master equation approaches. We investigate these features with four examples  which have entirely different physical origins  (i) a two qubit Werner state map with time dependent noise parameter (ii) Phenomenological model of a recent optical experiment (Nature Physics, {\bf 7}, 931 (2011)) on the open system evolution of photon polarization. (iii) Hamiltonian dynamics of a qubit coupled to a bath of $N$ qubits   and (iv) two qubit unitary dynamics of Jordan et. al. (\pra {\bf 70}, 052110 (2004) with initial product states of qubits. In all these models, it is shown that the positivity/negativity of the eigenvalues of intermediate time dynamical $B$ map determines the Markov/non-Markov nature of the dynamics.       
\end{abstract}
\maketitle

\section{Introduction}

Understanding the basic nature of  dynamical evolution of a quantum system, which interacts with an inaccessible environment, attracts growing importance in recent years~\cite{Breuer, Alicki}. This offers the key to achieve control over quantum systems  -- towards their applications in the emerging field of quantum computation and communication~\cite{Niel}. While the overall system-environment state evolves unitarily, the dynamics governing the system is described by a completely positive (CP), trace preserving  map~\cite{ECGS1, Choi,ji}.

Markov approximation holds when the future dynamics  depends only on the present state    -- and not on the history of the system i.e.,  memory effects are negligible. The corresponding Markov  dynamical map constitutes a trace preserving, CP, continuous one-parameter quantum  semi-group~\cite{Lindblad,GKS}.  Markov dynamics governing the evolution of the system density matrix is conventionally described by Lindblad-Gorini-Kossakowski-Sudarshan (LGKS)  master equation~\cite{Lindblad,GKS} $\frac{d\rho}{dt}={\cal L}\rho$, where ${\cal L}$ is the time-independent Lindbladian operator generating the underlying  quantum  Markov semi-group. Generalized  Markov processes are formulated in terms of time-dependent Lindblad generators and the associated trace preserving CP dynamical map is a two-parameter divisible map~\cite{div,RHP}, which  too corresponds to memory-less Markovian evolution. 

Not completely positive (NCP)  maps do make their presence felt in the open-system dynamics obtained from the
joint unitary  evolution --  if the system and environment are in an
initially quantum correlated state~\cite{Jordan,Rosario, Anil, Kavan}. In such cases, the open-system evolution turns out to be non-Markovian~\cite{AKRUS}. However, the source of such non-Markovianity could not be attributed entirely to either initial system-environment correlations or their dynamical interaction or both. This issue gets refined if initial global state is in the tensor product form, in which case the sole cause of Markovianity/non-Markovianity could be attributed to dynamics alone. It is known that the time evolution of a subsystem from an initial tensor product form to two different later times, $t_1,t_2\ (t_2>t_1)$, are both CP. However the dynamics in the intermediate time steps between $t_1$ and $t_2$  need not be CP. The quantum stochastic $A$  and dynamical  $B$  maps  -- first introduced as a quantum extension  of classical stochastic dynamics --  by Sudarshan, Mathews, Rau and Jordan (SMRJ)~\cite{ECGS1} nearly five decades ago, offer an elegant approach  to explore Markovian/non-Markovian  nature of open system evolution. The interplay of $A$ and $B$ maps at intermediate times,  to bring out the Markov or non-Markov avataras of open system evolution, is established in this paper.  

To place these ideas succinctly,  there are three basic aspects in open system quantum dynamics:  (1) nature of dynamical interaction between the system and its environment, (2) role of initial correlations in system-environment state and (3) nature of dynamics at intermediate times.  Last few years have witnessed  intense efforts towards understanding these~\cite{Jordan, Rosario, Anil, Kavan, div, RHP, NM,AKRUR,AKRUS, BLP,  haikka, kos, Hou}. The third issue is the focus here to discern  the Markov/non-Markov nature of dynamics in terms of intermediate time $A$ and $B$ maps.

The contents are organized as follows: In Sec.~II some basic concepts~\cite{ECGS1} on  $A$ and $B$ maps are given. The emergence of CP/NCP maps, at intermediate times, under open system dynamics is discussed in Sec.~III. Sec.IV is devoted to a powerful link (brought out by Jamilkowski isomorphism) between the $B$ map and the dynamical state. Some illustrative examples of dynamical $B$ map to investigate the CP/NCP nature of  dynamics at intermediate times  are discussed in Sec.~V. 
The examples are chosen from different origins: one based entirely from the general considerations of Jamiolkowski isomorphism; second one on the recent all-optical open system experiment to drive Markovian to non-Markovian transitions; the other two examples are based on open system Hamiltonian dynamics. In all these four examples, no master equation is employed in the deduction of Markov to non-Markov transitions --  but the CP/NCP nature of the intermediate dynamical map (via the sign of the eigenvalue of the $B$ map) has been invoked.  Sec.~VI has some concluding remarks.  

\section{Preliminary ideas on dynamical $A$ and $B$ maps}
The stochastic $A$ and  dynamical $B$  maps~\cite{ECGS1} transform the initial system density matrix $\rho_S(t_0)$ to final density matrix $\rho_S(t)$ via,
\begin{eqnarray}
\label{A}
\left[\rho_S(t)\right]_{b_1b_2}&=&\sum_{a_1,a_2} \left[A(t,t_0)\right]_{b_1b_2;a_1a_2}\, \left[\rho_S(t_0)\right]_{a_1a_2},  \\ 
\label{B}
\left[\rho_S(t)\right]_{b_1b_2}&=&\sum_{a_1,a_2} \left[B(t,t_0)\right]_{b_1a_1;b_2a_2}\, \left[\rho_S(t_0)\right]_{a_1a_2}, \\  && \ \ \ \  \ a_1,a_2,b_1,b_2=1,2,\ldots, d  \nonumber
\end{eqnarray} 
where the realligned matrix $B$ is defined by, 
\begin{equation}
\label{bdef}
B_{b_1a_1;b_2a_2}= A_{b_1b_2;a_1a_2}. 
\end{equation} 
The requirement that the evolved density matrix $\rho_S(t)$  has unit trace and  is Hermitian, positive semi-definite places the following conditions on $A$ and $B$~\cite{ECGS1}: 
{\scriptsize
\begin{eqnarray}
\label{prop}
{\rm Trace\ preservation}:&&\ \ \sum_{b_1}A_{b_1b_1;a_1a_2}=\delta_{a_1a_2}, \nonumber \\ 
&& \ \ \sum_{b_1}B_{b_1a_1;b_1a_2}=\delta_{a_1a_2} ,\nonumber \\
{\rm Hermiticity}:\ \ \ &&\ \ A_{b_1b_2;a_1a_2}=A^{*}_{b_2b_1;a_2a_1},\nonumber \\ 
&&  \ \ B_{b_1a_1;b_2a_2}=B^{*}_{b_2a_2;b_1a_1} \\
{\rm Positivity}: \ \ \  && \ \ \sum_{a_1,a_2,b_1,b_2}\,x^*_{b_1}\,x_{b_2}\, A_{b_1b_2;a_1a_2}\, y_{a_1}\,y_{a_2}^*\geq 0,\nonumber \\
&&  \ \ \sum_{a_1,a_2,b_1,b_2}\, x^*_{b_1}\, y_{a_1}\, B_{b_1a_1;b_2a_2}\,x_{b_2}\, y_{a_2}^*\geq 0 \nonumber 
\end{eqnarray}}
It may be readily identified that the dynamical $B$ map is positive, Hermitian $d^2\times d^2$ matrix with trace $d$ -- corresponding to CP evolution. We would also like to point out here that the composition of two 
stochastic $A$-maps, $A_1\ast A_2$ tranforming $\rho_S(t_0)\stackrel{A_1}{\longrightarrow}\rho_S(t_1)\stackrel{A_2}{\longrightarrow}\rho_S(t_2)$ is merely a matrix multiplication, whereas it is not so in its $B$-form. 

\section{CP/NCP nature of Intermediate time  $A$ and $B$ maps}  
Let us consider unitary evolution of global system-environment state $\rho_S(t_0)\otimes \rho_E(t_0)$ from an initial time $t_0$ to a final time $t_2$ -- passing through an intermediate instant $t_1$ (i.e., $t_0<t_1<t_2$). The $A$-map associated with $t_0$ to $t_1$ and that between $t_0$ to $t_2$ are  identified as follows:
\begin{eqnarray}
\label{a1}
&&{\rm Tr}_E\left[U(t_j,t_0)\rho_S(t_0)\otimes \rho_E(t_0)U^\dag(t_j,t_0)\right]=A(t_j,t_0)\, \rho_S(t_0)\nonumber \\ 
&&\ \ \ \ \ \ \ \ \ \ = \rho_S(t_j),\ \ j=1,2.  
\end{eqnarray} 
The stochastic map $A(t_j,t_0)$ is completely positive (correspondingly the dynamcal $B(t_j,t_0)$ matrix  is positive). 

In order to identify  the intermediate stochastic map $A(t_2,t_1)$, we make use of the composition law of unitary evolution $U(t_2,t_0)=U(t_2,t_1)U(t_1,t_0)$: 
{\scriptsize
\begin{eqnarray}
 \label{a2}
&{\rm Tr}_E\left[U(t_2,t_1)\left\{U(t_1,t_0)\rho_S(t_0)\otimes \rho_E(t_0)U^\dag(t_1,t_0)\right\}U^\dag(t_2,t_1)\right]=\nonumber \\
&\hskip 1in \ \    A(t_2,t_0)\, \rho_S(t_0).  
\end{eqnarray}} 
However, this  does not lead naturally to   $A(t_2,t_0)=A(t_2,t_1)A(t_1,t_0)$ for the $A$ map.   Invoking Markovian approximation (memory-less reservoir condition~\cite{note}) $\left\{U(t_1,t_0)\rho_S(t_0)\otimes \rho_E(t_0)U^\dag(t_1,t_0)\right\}\approx\rho_S(t_1)\otimes\rho_E(t_1)$,  
the LHS of (\ref{a2}) may be expressed as, 
\begin{equation}
 \label{a3}
{\rm Tr}_E\left[U(t_2,t_1)\rho_S(t_1)\otimes\rho_E(t_1)U^\dag(t_2,t_1)\right]=A(t_2,t_1)\, \rho_S(t_1).  
\end{equation} 
Further, substituting $j=1$ in (\ref{a1}) and expressing $\rho_S(t_0)=A^{-1}(t_1,t_0)\rho_S(t_1)$ in (\ref{a2}) the intermediate $A$ map $A(t_2,t_1)$ is identified: 
\begin{equation}
\label{a4}
A(t_2,t_1)=A(t_2,t_0)A^{-1}(t_1,t_0). 
\end{equation}  
In other words, when the environment is passive ( Markovian dynamics), the intermediate $A$-map has the divisible composition as in (\ref{a4}). In such cases $A(t_1,t_2)$ is ensured to be CP -- otherwise it is NCP, and hence non-Markovian. Correspondingly, the intermediate $B$-map $B(t_2,t_1)$ is positive if the dynamics is Markovian; negative eigenvalues of $B(t_2,t_1)$  imply non-Markovianity. 

\section{The $B$ map and the Jamiolkowski Isomorphism}

The Jamiolkowski isomorphism~\cite{ji} provides an insight  that the $B$-map is directly related to a $d^2\times d^2$ system-ancilla bipartite density matrix. More specifically, 
the action of the map $A^{\rm Id}\otimes A$ on the maximally entangled system-ancilla state $\vert\psi_{\rm ME}\rangle=\frac{1}{\sqrt{d}}\sum_{i=0}^{d-1}\vert i,i\rangle$ results in the density matrix $\rho_{ab}$ which may be identified to be related to the dynamical $B$ map i.e.,
\begin{equation}
\label{th}
\rho_{ab}=\left[A^{\rm Id}\otimes A \right]\vert\psi_{\rm ME}\rangle\langle\psi_{\rm ME}\vert\rightarrow \frac{1}{d}\, B 
\end{equation}
gives an explicit matrix representation for the $B$-map
(Here $A^{\rm Id}$ is the identity A-map, which leaves the ancilla undisturbed). 

In detail, we have,
 \begin{widetext}
 \begin{eqnarray}
\label{Jam}
\left(\rho_{ab}\right)_{a_1b_1;a_2b_2}&=&\sum_{a'_1,a'_2,b'_1,b'_2}\,\left[A^{\rm Id}\otimes A \right]_{a_1b_1a_2b_2;a'_1b'_1a'_2b'_2}\, 
\left[\vert\psi_{\rm ME}\rangle\langle\psi_{\rm ME}\vert\right]_{a_1'b_1';a_2'b_2'} \nonumber \\
&=& \frac{1}{d}\, \sum_{a'_1,a'_2,b'_1,b'_2}\, \delta_{a_1,a'_1}\delta_{a_2,a'_2}\, A_{b_1b_2;b_1'b_2'}\, \delta_{a_1',b_1'}\delta_{a_2',b_2'}\nonumber \\ 
&=& \frac{1}{d}\,  A_{b_1b_2;a_1a_2} = \frac{1}{d}\,  B_{b_1a_1;b_2a_2},
\end{eqnarray}
\end{widetext}
or $\left(\rho_{ba}\right)_{b_1a_1;b_2a_2}=\frac{1}{d}\,  B_{b_1a_1;b_2a_2}$.

In other words, Jamiolkowski isomorphism maps {\em every completely positive dynamical map} $B$ acting on $d$ dimensional space to 
a positive definite $d^2\times d^2$ bipartite density matrix $\rho_{ab}$ (See Eq.~(\ref{Jam}))   -- whose partial trace (over the first subsystem -- as seen  from the trace preservation property on dynamical map $B$ (as in Eq.(\ref{prop})) is a maximally disordered state.  One such set of bipartite $d\times d$ density matrices  belong to the class  that are invariant under $U \otimes U$~\cite{Vol}  -- which constitute the well-known Werner density matrices. One may now identify several toy models of dynamical $B$ maps -- including the two qubit Werner state example motivated by the above remark  --  to investigate the nature of intermediate time dynamics. 

In the next section we present specific examples chosen to illustrate the features of intermediate dynamical maps: (i) A toy model map inspired by Jamiolkowski isomorphism -- which is not based on any Hamiltonian underpinning. (ii) Recent optical experiment by Liu et. al.,~\cite{Liu} on open system evolution of photon polarization  to bring out non-Markovianity features is reinterpreted in terms of NCP nature of the intermediate dynamical map.  (iii) Intermediate dynamical map in the Hamiltonian evolution of a two-level system coupled to  $N$ two-level systems~\cite{Hou} (iv) open system dynamics arising from a two qubit  unitary evolution ~\cite{Jordan}.     

\section{Examples} 

\subsection{A toy model dynamical map}

The two qubit Werner density matrix is a natural choice for a prototype of  dynamical $B$-map -- arising from general considerations of the Jamiolkowoski isomorphism:  
\begin{equation}
B(t,0)=\frac{[1-p(t)]}{2}\, I_2\otimes I_2 +\frac{p(t)}{2}\, \vert\Psi^{(-)}\rangle\langle\Psi^{(-)}\vert
\end{equation}  
 with a time dependent noise parameter $0\leq p(t) \leq 1$, and $\vert\Psi^{(-)}\rangle=\frac{1}{\sqrt{2}}\left(\vert 0, 0\rangle-\vert 1,1\rangle\right)$ is the Bell state. For a dynamical map, time dependence in $p(t)$ occurs due to the  underlying Hamiltonian evolution.   This state  
 is especially important in that it exhibits both separable and entangled states, as its characteristic parameter $p(t)$ is varied. Its use here as a valid $B$-map is novel in identifying transitions between Markovianity and non-Markovianity in the dynamics as captured from their intermediate time behavior.

On evaluating the corresponding $A$ map $A(t,0)$ (expressed in the standard $\{\vert 0,0\rangle, \vert 0,1\rangle, \vert 1,0\rangle, \vert 1,1\rangle\}$ basis) i.e.,  
\begin{equation} 
A(t,0)=\left(\begin{array}{cccc} \frac{1+p(t)}{2} & 0 & 0 & 0 \\ 
                              0 & \frac{1-p(t)}{2} & -\frac{p(t)}{4} & 0 \\ 
                              0 & -\frac{p(t)}{4} & \frac{1-p(t)}{2} & 0 \\ 
                              0 & 0 & 0 & \frac{1+p(t)}{2}      \end{array}\right) \nonumber \\ 
\end{equation}
one can obtain the intermediate dynamical map 
$A(t_2,\,t_1)=A(t_2,\,0)A^{-1}(t_1,\,0)$. The intermediate time $B$-map $B(t_2,t_1)$ is given by 
\be
B(t_2,\,t_1)=\frac{1}{2}\left(1-\frac{p(t_2)}{p(t_1)}\right)\, I_2\otimes I_2+\frac{2p(t_2)}{p(t_1)}\, \vert\Psi^{(-)}\rangle\langle\Psi^{(-)}\vert.
\ee
Its eigenvalues are $\lambda_1=\lambda_2=\lambda_3=\frac{1}{2}\left(1-\frac{p(t_2)}{p(t_1)}\right)$ and $\lambda_4=\frac{1}{2}\left(1+\frac{3\, p(t_2)}{p(t_1)}\right)$.  

A choice $p(t)=\cos^{2M}(a t)$ for any $M\geq 1$ leads to NCPness of the intermdiate map  -- as  the eigenvalues   $\lambda_{1,\,2,\,3}\equiv\lambda$ of $B(t_2,t_1)$  may assume  negative values -- and hence  non-Markovian  dynamics ensues. We have plotted the negative eigenvalue $\lambda$ of $B(t_2,t_1)$ as a function of  $\mu=t_2/t_1$ and  for typical values of $M=1,3,5$ in Fig.~1. This reveals transitions from  Markovianity to non-Markovianity and back in this model. 
\begin{figure}[h]
\includegraphics*[width=2.5in,keepaspectratio]{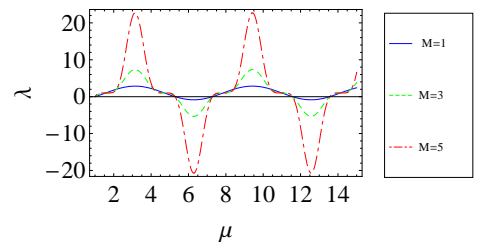}
\caption{A plot of the eigenvalue $\lambda$ of $B(t_2,\, t_1)$ versus $\mu=\frac{t_2}{t_1}$ for different values of $M$. The dynamics is non-Markovian when $\lambda$ assumes negative values and otherwise it is Markovian.}   
\end{figure}

Another choice  $p(t)=e^{-\alpha t}$ corresponds to a CP intermediate map -- resulting entirely in a Markovian process. In this case, we also find that $A(t_2,t_1)=A(t_2-t_1)$ and this forms a Markov semigroup. 
However, if $p(t)=e^{-\alpha\, t^\beta},\ \ (\beta\neq 1)$, the intermediate map is still CP (and hence Markovian),  though  $A(t_2,t_1)\neq A(t_2-t_1)$ and therefore,  it does not constitute a one-parameter semigroup.   

Furthermore,  we wish to illustrate through this toy model that concurrence  of $\rho_{ab}(t)=\frac{1}{d} \, B(t,0)$ 
(given by $C=\frac{3p(t)-1}{2}$) can never increase as a result of  Markovian evolution. This is because ensuing dynamics is a  local  CP map on the system. Any temporary regain of system-ancilla entanglement during the course of evolution is clearly attributed to the back-flow from environment to the system -- which is a signature of non-Markovian process. This feature is displayed in Fig.~2 by plotting the concurrence of $\rho_{ab}(t)$ for different choices of $p(t)$.  
\begin{figure}[ht]
\includegraphics*[width=2.5in,keepaspectratio]{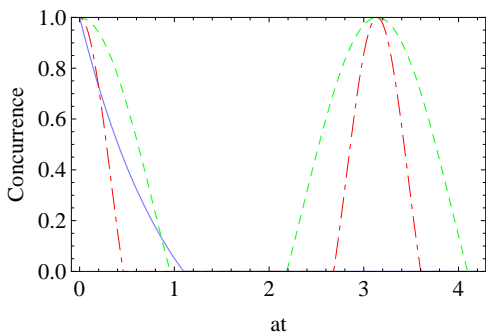}
\caption{Concurrence $C=\frac{3p(t)-1}{2}$ of the system-ancilla state $\rho_{ab}(t)=\frac{[1-p(t)]}{4}\, I_2\otimes I_2 +p(t)\, \vert\psi_{\rm ME}\rangle\langle\psi_{\rm ME}\vert$, vs scaled time $a\,t$, for the following choices (i) Markov process: $p(t)=e^{-at}$ (solid line) and (ii) non-Markov process:  $p(t)=\cos^{2M}(a\, t),\ M=1$ (dashed line) and $M=5$ (dot-dashed line). Note that there is a death and re-birth of entanglement (dash, dot-dashed lines) due of  back-flow from environment.}   
\end{figure} 
\subsection{Optical Experiment}

Recently, Liu et al~\cite{Liu} reported an optical experiment on the open quantum system constituted by the  
polarization degree of freedom of photons (system)  coupled to the frequency
degree of freedom (environment). They reported transition between Markovian and
non-Markovian regimes. 

The dynamical evolution of the horizontal and vertical poloarization states $\vert H\rangle, \vert V\rangle$ of the photon is captured by the following transformation: 
\begin{eqnarray}
\vert H\rangle \langle H\rangle &\mapsto&  \vert H\rangle \langle H\vert  \nonumber \\
\vert V\rangle \langle V\rangle &\mapsto&  \vert V\rangle \langle V\vert  \nonumber \\
\vert H\rangle \langle V\rangle &\mapsto&  \kappa^*(t)\, \vert H\rangle \langle V\vert \\
\vert V\rangle \langle H\rangle &\mapsto&  \kappa(t)\, \vert V\rangle \langle H\vert \nonumber 
\end{eqnarray}
where $\kappa(t)$ denotes the decoherence function, magnitude of which is modelled as (for details see \cite{Liu}), 
\begin{eqnarray} 
\vert \kappa(t)\vert&=&e^{-\frac{1}{2}\sigma^2\tau^2}\sqrt{1-4 A_1(1-A_1)\, \sin^2(\tau\, \Delta \omega)} \\ 
&& \hskip 1in 0\leq A_1\leq 1. \nonumber
\end{eqnarray}
The corresponding $A$ and $B$ maps (in the $\{HH, HV, VH, VV\}$ basis) are readility identified to be, 
\begin{eqnarray} 
A(t,0)&=&\left(\begin{array}{cccc} 1 & 0 & 0 & 0 \\ 
                              0 & \kappa^*(t) &0 & 0 \\ 
                              0 & 0 & \kappa(t) & 0 \\ 
                              0 & 0 & 0 & 1      \end{array}\right) \nonumber \\ 
B(t,0)&=&\left(\begin{array}{cccc} 1 & 0 & 0 & \kappa^*(t) \\ 
                              0 & 0 &0 & 0 \\ 
                              0 & 0 & 0 & 0 \\ 
                              \kappa(t) & 0 & 0 & 1      \end{array}\right)                                
\end{eqnarray}   
We construct the intermediate time dynamical map $B(t_2,t_1)$ from the corresponding $A(t_2,t_1)=A(t_2,0)A^{-1}(t_1,0)$ to obtain, 
\begin{equation}
 B(t_2,t_1)=\left(\begin{array}{cccc} 1 & 0 & 0 & \frac{\kappa^*(t_2)}{\kappa*(t_1)} \\ 
                              0 & 0 &0 & 0 \\ 
                              0 & 0 & 0 & 0 \\ 
                              \frac{\kappa(t_2)}{\kappa(t_1)} & 0 & 0 & 1      \end{array}\right).  
\end{equation}   
Eigenvalues of $B(t_2,t_1)$ are given by, 
\begin{equation}
\lambda_{1,4}= 1\pm \left\vert\frac{ \kappa(t_2)}{\kappa(t_1)}\right\vert, \lambda_{2,3}=0.  
\end{equation} 
The eigenvalue $\lambda_4$ can assume negative values indicating Markovian/non-Markovian regimes. A plot of the negative eigenvalue as a function of $A_1$, for different ratios $t_2/t_1$,  is given in Fig.~3  -- where one can clearly see the Markovian ($\lambda_4\geq 0$) and non-Markovian ($\lambda_4< 0$) regimes in this model.  
\begin{figure}[h]
\includegraphics*[width=2.5in,keepaspectratio]{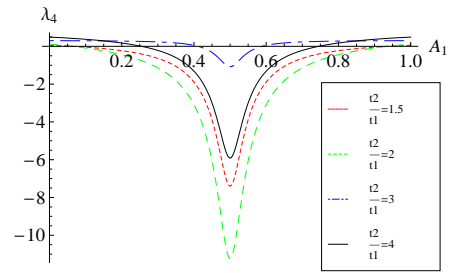}
\caption{A plot of the eigenvalue $\lambda_4$ of $B(t_2,\, t_1)$ versus $A_1$ for different values of $\mu=\frac{t_2}{t_1}$.}
\end{figure}
\subsection{Hamiltonian evolution of a two level system coupled to a bath of $N$ spins} 
We now present a  Hamiltonian model, which give rise to explicit structure of time dependence in the open system evolution.  Interaction Hamiltonian considered here is~\cite{Hou} 
\be
H=\frac{A}{\sqrt{N}}\sigma_z \sum_{k=1}^N \, \sigma_{k\,z}.
\ee                             
This is a simplified model of a hyperfine interaction of a spin-1/2 system with $N$ spin-1/2 nuclear environment in a quantum dot.  
Taking the initial system-environment state to be $\rho_S(0)\otimes \frac{I_{2^N}}{2^N}$, the dynamical $A$-map is obtained by evaluating 
${\rm Tr}_E\left[U(t,0)\,\rho_S(0)\otimes \frac{I_{2^N}}{2^N}\, U^\dag(t,0)\right]$ (where $U(t,0)={\rm Exp}[-i\,H\, t]$): 
\begin{eqnarray}
A(t,0)&=&\frac{1}{2}\left(1-x(t)\right)\sigma_z\otimes\sigma_z+\frac{1}{2}\left(1+x(t)\right)I_2\otimes I_2,  \nonumber \\
& & x(t)=\cos^N \left (\frac{2At}{\sqrt N}) \right). 
\end{eqnarray}
From this, the intermediate map $A(t_2,t_1)$ (see (\ref{a4})) and in turn the corresponding $B(t_2,t_1)$ may be readily evaluated. We obtain, 
\begin{eqnarray} 
B(t_2,\,t_1)&=&\frac{1}{2}\, \left(I_2\otimes I_2+\sigma_z\otimes\sigma_z\right)\nonumber \\ && +\frac{x(t_2)}{2x(t_1)}\left(\sigma_x\otimes\sigma_x-\sigma_y\otimes\sigma_y\right).
\end{eqnarray}  
The eigenvalues of $B(t_2,t_1)$ are $0,0,1\pm\frac{x(t_2)}{x(t_1)}.$ Clearly, the intermediate time dynamics exhibits NCP as one of the eigenvalues i.e., 
$\lambda=1-\frac{x(t_2)}{x(t_1)}$ 
of $B(t_2,t_1)$ can assume negative values. We illustrate regimes of Markovianity/non-Markovianity revealed via positive/negative values of $\lambda$ (plotted as a function of $\mu=t_2/t_1$) in Fig.~4.   
\begin{figure}[h]
\includegraphics*[width=2.5in,keepaspectratio]{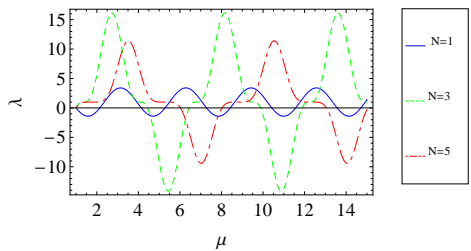}
\caption{The variation of the eigenvalue $\lambda$ of $B(t_2,\, t_1)$ (as a function of $\mu=t_2/t_1$) from positive to negative values and back with the passage of time for different values of $N$.}   
\end{figure}
\subsection{Two qubit unitary evolution} 
We now consider the open system dynamics arising from the  unitary evolution~\cite{Jordan} 
\begin{eqnarray}
U(t,0)&=&e^{-i\,t\, [\omega\, \sigma_{z}\otimes \sigma_{x}]} \\
&=&\cos(\omega\, t/2)\, I_2\otimes I_2-i\sin(\omega\ t/2)\, \sigma_z\otimes \sigma_x \nonumber
\end{eqnarray}
on the system-environment initial state $\rho_{SE}(0)=\rho_S(0)\otimes \rho_E(0)=\frac{1}{2}\left(I_2+\sigma_x\right)\otimes\frac{1}{2}\left(I_2+\sigma_z\right).$ The $A(t,0)$ map is given by, 

\be
A(t,0)=\frac{1}{2}\left(1+\cos(\omega\, t)\right)\, I_2\otimes I_2+\frac{1}{2}\left(1-\cos(\omega\, t)\right)\, \sigma_z\otimes\sigma_z. 
\ee
Following (\ref{a4}), we obtain
\begin{eqnarray}
B(t_2,t_1)&=&\frac{1}{2}\, \left(I_2\otimes I_2+\sigma_z\otimes\sigma_z\right)\nonumber \\ && +\frac{\cos(\omega\,t_2)}{2\cos(\omega t_1)}\left(\sigma_x\otimes\sigma_x-\sigma_y\otimes\sigma_y\right).
\end{eqnarray}  
The eigenvalues of the $B$-map are given by $0,\ 0, 1\pm \left\vert\frac{\cos \omega t_2}{\cos \omega t_1}\right\vert$. The eigenvalue $\lambda=1\pm \left\vert\frac{\cos \omega t_2}{\cos \omega t_1}\right\vert$ can assume negative values -- bringing out the non-Markovian features prevalent in the dynamical process. Fig.~5 illustrates the transitions from Markovianity to non-Markovianity. 
\begin{figure}[h]
\includegraphics*[width=2.5in,keepaspectratio]{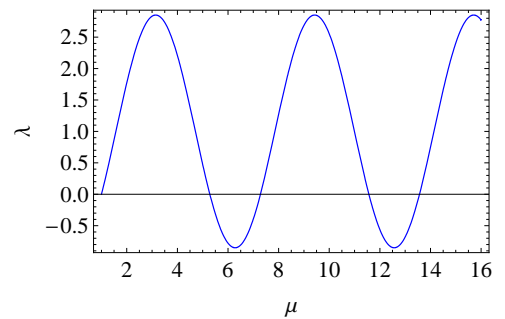}
\caption{The plot of the eigenvalue $\lambda=1- \left\vert\frac{\cos \omega t_2}{\cos \omega t_1}\right\vert$ as a function of  $\mu=\frac{t_2}{t_1}$. The periodic transitions of $\lambda$ from positive to negative values indicates the transition of the process from Markovian to non-Markovian.}  
\end{figure} 
This model, with initially correlated states,  has been explored before in Refs.~\cite{Jordan, AKRUS} and the dynamical map turned out to be NCP throughout not merely in the intermediate time interval).    
\section{Summary}

In conclusion, a few remarks on a 
variety of definitions of non-Markovianity in the recent literature may be recalled here.  Mainly the focus has been towards capturing the violation of semi-group property~\cite{AKRUR,AKRUS} or  more recently -- its two-parameter generalization viz the divisibility of the dynamical map~\cite{div,RHP}. Yet another measure, where non-Markovianity~\cite{BLP} is attributed to  increase of distinguishability of any pairs of states (as a result of the partial back-flow of information from the environment  into the system) and is quantified in terms of trace distance of the states. It has been shown that the two different measures of non-Markovianity  -- one based on the divisibility of the dynamical map~\cite{RHP}  and the other based upon the distinguishability of quantum states~\cite{BLP} -- need not agree with each other~\cite{haikka}.    A modified version of the criterion of Ref.~\cite{RHP} was proposed recently~\cite{Hou}. In this paper we have established the interplay of stochastic $A$  and dynamical $B$ maps at intermediate times, revealing Markovian/non-Markovian regimes. We have explored four different examples  revealing the features of intermediate time maps originating from variety of physical mechanisms : (i) A toy model map  inspired by general considerations based on  Jamiolkowski isomorphism -- which explores a two qubit Werner state  with  time-dependent noise parameter as a dynamical map (ii) A reinterpretation of the phenomenological model explaining the recent optical experiment by Liu et. al.,~\cite{Liu} in terms of NCP nature of the intermediate $B$ map.  (iii) Hamiltonian evolution describing the hyperfine interaction of a spin-1/2 system with $N$ spin-1/2 nuclear environment in a quantum dot~\cite{Hou} displaying Markovian/non-Markovian behaviour and (iv) Unitary evolution of Jordan et. al., ~\cite{Jordan} -- wherein initial system-environment two qubit is chosen in a product state. Here too, intermediate time dynamical map exhibits Markov/non-Markov regimes. It is interesting to note that   the dynamics had been identified to be NCP throughout not merely in the intermediate time interval -- when initially correlated states were employed~\cite{Jordan, AKRUS}. Placing these two results together, brings forth that the source of non-Markovianity in this model is  attributable entirely to the unitary dynamics --- rather than initial correlations of system-environment qubits. We have thus exposed the underlying features of intermediate time $A$ and $B$ maps to bring out clearly if the dynamics relies on past history of the states or not.


\end{document}